\documentstyle[prd,aps]{revtex}
\begin{document}

\draft
\preprint{{}}
\title{Supersymmetric Left-Right Models
and Light Doubly Charged Higgs Bosons and Higgsinos}
\author{Z. Chacko and R. N. Mohapatra }
\address{ Department of Physics, University of Maryland,
College Park, MD-20742, USA.}
\date{December, 1997}
\maketitle
\begin{abstract}
We point out that in a large class of supersymmetric left-right models
with automatic R-parity conservation there are a pair of light doubly
charged Higgs bosons and Higgsinos. Requiring the mass of these particles 
to satisfy the LEP Z-width bound implies that $W_R$ mass must be
above $10^{9}$ GeV. 
\end{abstract} \pacs{\hskip 6 cm  UMD-PP-98-67}

\vskip2pc

\section{Introduction}

Supersymmetric left-right models (SUSYLR) are attractive for several reasons:
1) they imply automatic conservation of baryon and lepton 
number\cite{moh}, a property
   which makes the standard model so attractive, but is not shared by the
   minimal supersymmetric standard model (MSSM);
2) they provide a natural solution to the strong and weak CP problems of the
   MSSM\cite{rasin};
3) They yield a natural embedding of the see saw mechanism for small neutrino 
   masses\cite{gell};
4) they arise as an intermediate scale theory in several SUSY GUT's.
   
An essential feature of these models is that the
$SU(2)_R \times U(1)_{B-L}$ symmetry is broken down to $U(1)_Y$ by the
vacuum expectation values of a pair of Higgs multiplets which 
transform as the adjoint of the $SU(2)_R$ group with $B-L=\pm 2$.
One of these is the same Higgs multiplet (denoted by $\Delta^c$ below) 
which is 
used to implement the  see-saw mechanism for small neutrino masses.
Both of them contain doubly charged Higgs bosons and Higgsinos. It has 
recently been shown\cite{kuchi} that in simple versions of this theory 
where the hidden sector SUSY breaking scale is above the $M_{W_R}$, the 
ground state 
breaks R-parity unless higher dimensional nonrenormalizable terms are
included\cite{rasin,aulakh}.

It is the goal of this paper to show that the constraints of
supersymmetry imply that the above mentioned doubly charged particles 
are very light
in most simple and intersting versions of this model. Since these masses
depend on the scale ($v_R\equiv v$) of the $SU(2)_R$ breaking, one can use 
the LEP Z-width constraints to fix a lower bound on $v_R$.
It turns out that parity invariance does not play any role in our proof.
Therefore, the bound on $v_R$ applies also to models based on the gauge
group $SU(2)_L\times SU(2)_R\times U(1)_{B-L}$ without parity
as long as the $SU(2)_R$ symmetry is broken by the same type of fields. 

Furthermore, the existence of these light fields is independent of the 
scale at which supersymmetry 
is broken provided that the effect of the breaking on the Higgs sector is soft.
However the mass splitting between the Higgs bosons and Higgsinos is
crucially dependent on whether the scale of supersymmetry breaking is 
higher than the $W_R$ scale (as in a supergravity mediated scenario) or
lower than the $W_R$ scale (as in a gauge mediated scenario). This
causes the bounds on the $W_R$ scale to arise differently in the two
cases. In the former, the bound arises from considering the Higgs boson 
masses and in the latter case from the Higgsino masses. 

The underlying reason for the existence of these light particles can be 
understood by considering the result of reference\cite{kuchi}. The particle 
content
of this model consists of bidoublets $\phi$ (2,2,0), Higgs fields $\Delta$
(3,1,2), $\bar{\Delta}$(3,1,-2), $\Delta^c $(1,3, -2) and 
$\bar{\Delta ^c} $(1,3,+2) and a
singlet in addition to the usual quarks and leptons. (The numbers in 
parenthesis refer to their transformation properties under $ SU(2)_L \times 
SU(2)_R \times U(1)_{B-L}$ ). If only renormalizable interactions of these
fields are considered and R-parity is unbroken (i.e. $<\tilde{\nu^c}>= 
0$) then in the absence of 
supersymmetry breaking there exist a continuously connected set of vacua 
parametrised by a single angle $\theta$ such that
 \begin{eqnarray}
\Delta^c = 
\left[
\matrix{ {0} & {v  cos  \theta} \cr
         {v  sin  \theta} & {0} \cr}
\right]
\end{eqnarray}

\begin{eqnarray}
\bar {\Delta^c} = 
\left[
\matrix{ {0} & {v  sin  \theta} \cr
         {v  cos  \theta} & {0} \cr}
\right]
\end{eqnarray}

If $\theta=0$, electric charge is conserved; otherwise it is broken.Thus
the only phenomenologically viable vacuum, the charge conserving one, is 
degenerate with a continuously connected set of other vacua. The 
excitation that
corresponds to the flat direction connecting all these vacua must be a massless
particle; it is straightforward to verify that in the charge conserving vacuum
this particle is doubly charged. The other doubly charged particle is also
massless but does not correspond to a flat direction. Once supersymmetry is
broken the flat direction 
is lifted and the theory will choose to live either in the `good' vaccum or the
`bad' vacuum. The main result of reference\cite{kuchi} is that if the 
scale at which supersymmetry is broken is higher than the $W_R$ scale 
and electroweak symmetry breaking is ignored, then the theory 
necessarily lives in the charge violating vacuum if R-parity is unbroken.
However if non-renormalizable operators suppressed by powers of $M_{Planck}$
are added to the theory\cite{rasin}, then for sufficiently high right handed 
scale the 
theory can live in the charge preserving vacuum\cite{aulakh}. However, this 
suggests that
the previously massless Higgs boson corresponding to a flat direction is
still light, since the flat direction
has only been given a positive slope by these higher dimensional operators.
We calculate the
mass of this Higgs boson and verify that it is indeed light. We use the 
experimental lower bound on the mass of such a doubly charged particle to put
a lower bound on the $W_R$ scale when electroweak symmetry 
breaking is ignored. We then show that the inclusion of electroweak symmetry
breaking does not alter this result. Indications of such a lower bound in
a very specific model was also noted in Ref.\cite{aulakh}. Our result
is however more general.

If however the $W_R$ scale is above the supersymmetry breaking scale, as in  
gauge mediated supersymmetry breaking scenarios\cite{dine}, we show that 
even the 
renormalizable theory may live in the charge preserving vacuum. In this case
the light Higgs bosons pick up a mass from the breaking of supersymmetry
which is of the same order of magnitude as the masses of the 
superpartners of 
the standard model particles. Now however the corresponding Higgsinos are very
light since the breaking of supersymmetry is assumed to be soft. Thus the 
non-renormalizable operators are once again needed, this time to give mass
to the Higgsinos. The experimental lower bound on the mass of such particles
can once again be used to put a lower bound on the right handed scale.

Finally, we also point out that a light doubly charged Higgs and Higgsino
fields are
also present in the version of the model where the vacuum state breaks 
R-parity since it was shown in Ref.\cite{kuchi} that in these models 
there is an upper limit on the $W_R$ scale of order of a TeV and we 
expect the masses of all particles in the theory to be at most of the
order of the $W_R$ mass.

One may thus infer that the only consistent low $W_R$ scale theory is
the one that breaks R-parity dynamically.

Although our detailed analysis is limited to a specific class of models, we 
consider whether our result holds for models with further matter content.
We find that SUSYLR theories necessarily imply these light doubly charged
Higgs superfields unless the model contains exotic light doubly charged 
$SU(2)_R$
singlets or certain additional Higgs multiplets which break $SU(2)_R$ while
preserving hypercharge. Our results also have important implications for 
coupling constant unification in SUSY GUTs.

\section{Analysis for $W_R$ scale below SUSY breaking scale}

In this section we calculate the masses of the doubly charged Higgs bosons
and Higgsinos for a specific class of models when supersymmetry
is broken above the $W_R$ scale as in a supergravity mediated scenario. 
We first study the theory classically and
subsequently justify our conclusions where required using an effective field
theory analysis. 

The matter content of the model we examine consists of :
the quarks $Q$(2,1,1/3) and $Q^c$(1,2,-1/3); the leptons $L$(2,1,-1) and 
$L^c$(1,2,1); the electroweak Higgs bidoublet $\phi$(2,2,0);
the triplets $\Delta$(3,1,2),$\Delta^c$(1,3,-2),$\overline {\Delta}$
(3,1,-2) and $\overline {\Delta}^c$(1,3,2) and an arbitrary number of 
singlets $S_i$(0,0,0)
where the numbers in parenthesis refer to their transformation properties 
under $ SU(2)_L \times SU(2)_R \times U(1)_{B-L} $ respectively. In a later
section we will consider models with additional matter content. Our procedure
will be to first write
down the most general potential involving the above fields consistent 
with the
symmetries and from there to obtain the mass matrices of the doubly charged
fields and show that one of the eigenvalues of the Higgs boson mass matrix is
light.

We consider the most general superpotential consisting of the above fields.
In order to account for the possibility that right handed scale is large, we 
include, in addition
to the renormalizable interactions, all possible nonrenormalizable 
interactions of the $\Delta$'s and $\Delta^c$'s among themselves to 
lowest order in 
1/$M_{Planck}$. Then, the relevant part of the superpotential is

\begin{eqnarray}
W &=& ifL^{c^T}\tau_2\Delta^c L^c + ( M_0 +\lambda S_1) Tr(\Delta^c 
\overline \Delta^c) +
   G(S_i,X_i) + A [Tr(\Delta^c \overline {\Delta}^c)]^2/2 \nonumber\\
&& + B Tr(\Delta^c 
\Delta^c) Tr (\overline{\Delta}^c \overline{\Delta}^c)/2
\end{eqnarray}

In the above equation, $X_i$ is a generic label for any field apart from the 
$\Delta^c$'s 
that the  $S_i$'s couple to; A, B, f, $\lambda$ and M are parameters 
of the theory and $G(S_i,X_i)$ 
is the most general superpotential in the $S_i$'s and $X_i$'s alone. 
A and B are of order 1/$M_{Planck}$.

Using Eq. 3, one can give a group theoretical argument for the existence of 
light doubly charged particles in the supersymmetric limit as follows. For 
this purpose let us first
ignore the higher dimensional terms $A$ and $B$ as well as the leptonic 
couplings $f$. It is then clear that the superpotential has a 
complexified $U(3)$ symmetry (i.e. a $U(3)$ symmetry whose parameters are
taken to be complex) that operates on the $\Delta^c$ and $\bar\Delta^c$
fields. This is due to the holomorphy of the superpotential. After one 
component of each of the above fields acquires vev as in the charge 
conserving case with $\theta=0$ (and supersymmetry
guarantees that both vev's are parallel), the resulting symmetry is the
complexified $U(2)$. This leaves 10 massless fields. Once we bring in the
D-terms and switch on the gauge fields, six of these fields become massive
as a consequence of the Higgs mechanism of supersymmetric theories.
That leaves four massless fields in the absence of higher dimensional 
terms. These are the two complex
doubly charged fields. Of the two non-renormalizable terms $A$ and $B$,
only the A-term has the complexified $U(3)$ symmetry. Hence the 
supersymmetric contribution to the doubly charged particles will come only
from the B-term. Although the leptonic couplings do not respect this
symmetry, they are unimportant in determinimg the vacuum structure as 
long as R-parity is conserved and hence they do not alter our result. 
Let us now proceed to prove this via explicit calculations.  

The most general soft supersymmetry breaking
potential compatible with the symmetries and relevant for our analysis has 
the form

\begin{eqnarray}
V_S  &=& M_L^2 {L^c}^{\dagger}{L^c} +
M_1^2 Tr({\Delta^c}^{\dagger}\Delta^c)+ \nonumber\\
&& M_2^2 Tr( {\overline {\Delta}^c}^{\dagger}\overline {\Delta^c})
+ \Sigma_i \left(\lambda_i 'M_S Tr (\Delta ^c \bar {\Delta^c})S_i\right) + 
M'^2 Tr(\Delta^c \overline {\Delta^c}) \nonumber\\
&& + i f' M''{L^c}^T \tau_2\Delta^c L^c 
+ G'(S_i,X_i, S^{\dagger}_i,X^{\dagger}_i) + h.c. 
\end{eqnarray}       

where $M_L$, $M_1$, $M_2$, $M'$, $\lambda_i'$, $M_S$, f'and $M''$ are 
parameters
and $G'[S_i,X_i,S_i^{\dagger},X_i^{\dagger}]$ is an arbitrary function in 
the $S_i$'s, $X_i$'s and
their hermitian conjugates consistent with the soft breaking of SUSY.
The relevant part of the D terms have the form

\begin{eqnarray}
V_D &=&\frac{ g_2^2}{8} \Sigma_i (2Tr{\Delta^c}^{\dagger}\tau_i \Delta^c
 +2Tr {\overline{\Delta}^c}^{\dagger}\tau_i \overline {\Delta}^c 
  + Tr \phi^{\dagger}\tau_i\phi +L^{c\dagger}\tau_iL^c +{\bar 
L}^{c\dagger}\tau_i\bar L^c )^2 \nonumber\\
&& + \frac{g_1^2}{2} (Tr{\Delta^c}^{\dagger}\Delta^c 
 -Tr {\overline {\Delta}^c }^{\dagger}\overline {\Delta}^c -\frac{1}{4}
{L^{c\dagger}}L^c +\frac{1}{4}{\bar L}^{c\dagger}\bar L^c )^2
\end{eqnarray}


Since the masses of the doubly charged fields will be dependent on the 
neutral Higgs vevs, we first set out to determine these.
The potential to be minimised in order to determine the Higgs vevs consists 
of a sum of F,D and soft supersymmetry breaking terms. We assume that 
R-parity is unbroken so that  
$<\tilde L >$ = $<\tilde L^c>$ = 0 and the terms involving these fields are 
unimportant in determining the classical vacuum. Then the potential to
determine the Higgs vevs can be written as

\begin{eqnarray}
V &=& [ K {\overline {\Delta}^c}^i_j + A~Tr(\Delta^c \overline {\Delta}^c)
{\overline {\Delta}^c}^i_j + B~ Tr (\overline {\Delta}^c \overline{\Delta}^c)
{{\Delta}^c}^i_j][ h.c.] \nonumber\\
&& + [ K {{\Delta}^c}^i_j + A~ Tr(\Delta^c {\bar \Delta}^c)
{{\Delta}^c} ^i_j + B Tr (\overline {\Delta}^c \overline {\Delta}^c)
{\overline {\Delta}^c} ^i_j][ h.c.] \nonumber\\
&& + \beta [Tr(\Delta^c \overline {\Delta}^c)][ h.c.] 
+[\alpha_0 Tr(\Delta^c \overline {\Delta}^c) + h.c.] \nonumber\\
&& + M_1^2 Tr({\Delta^c}^{\dagger}\Delta^c)
+ M_2^2 Tr( {\overline {\Delta}^c}^{\dagger}\overline {\Delta}^c)
+ V_D
\end{eqnarray}

Here $\beta$, $\alpha_0$, and $K$ are parameters; the latter two in general
depend upon the expectation value of the singlet and its F-component. 
While $K = \lambda <S_1> + M $, $\alpha_0$ picks up contributions from the 
F component of $S_1$ in the form
$<\partial G / \partial{S_1}>^{\dagger}[Tr(\Delta^c \overline {\Delta}^c)]$
and also from $M'^2$ and $\lambda_iM_S$ in the soft supersymmetry breaking 
terms. $\beta$ is simply $\lambda ^2$.

We now assume the pattern of symmetry breaking to be such as to break 
$SU(2)_R$ while preserving electric charge and R-parity. As has been shown
in Ref.\cite{kuchi,aulakh}, for the class of models we are interested in,
this is always true. 
In an arbitrary SUSYLR model of this class it is
certainly not guaranteed that the pattern of left-right symmetry breaking
will work out correctly. However what we intend to show is that for any
model of this type which does have the right pattern of left-right 
symmetry breaking
and lives in the charge preserving vacuum the doubly charged fields 
will be light. 
If our assumption that the theory lives in the `good' vacuum is true 
then the masses of the doubly charged 
Higgs bosons will turn out to be positive at the end of our calculation;
 if not at least one of
these fields will turn out to have negative mass thereby giving us a
consistency check on our assumption. 
In order to give our proof, let us expand the Higgs field in components as 
 
\begin{eqnarray}
\Delta^c = 
\left[\begin{array}{cc}
 {\Delta^c}^{-}/\sqrt{2} & {v + \eta} \\           
 
         {\Delta^c}^{--} & -{\Delta^c}^-/\sqrt{2} \end{array}
\right]
\end{eqnarray}

\begin{eqnarray}
\overline {\Delta^c} = 
\left[
\matrix{ {\overline {\Delta}^c}^+/\sqrt{2} & {\overline {\Delta}^c}^{++} \cr
  \bar v +\bar \eta & -{\overline {\Delta}^c}^+/\sqrt{2} \cr} \right]
\end{eqnarray}

while the electroweak Higgs bidoublet $\phi$ has the vevs

\begin{eqnarray}
\phi = 
\left[
\matrix{ \kappa_u & 0 \cr
         0 & \kappa_d \cr}
\right]
\end{eqnarray}

While $M_1^2, M_2^2$ and $\beta$ are necessarily real, $K, \alpha_0,
A$ and $B$ are in general complex. By redefining the ${\Delta^c}$'s
by phase factors we can make $v$ and $\bar v$ real.
Defining $\alpha = Re[\alpha_0]$ and $\alpha'= Im[\alpha_0]$
this implies that the imaginary part of $\alpha_0$ satisfies 
the equation

\begin{equation}
\alpha' v\bar{v} + Im[Av\bar{v}(K+Av\bar{v})^*]=0
\end{equation}

The potential to be minimised in order to determine $v$,$\bar v$ is

\begin{eqnarray}
V &=& \beta v^2 \bar v^2 + 2 \alpha v \bar v + 
(K + A v \bar v)(K + A v \bar v)^* ({\bar v}^2
    + v^2) \nonumber\\
    && + M_1^2 v^2 + M_2^2 {\bar v}^2 + \frac{g_1^2}{2}[{ v^2-{\bar 
v}^2}]^2 +
    \frac{g_2^2}{8} [ 2({v^2-{\bar v}^2) + {\kappa_U}^2 -{\kappa_D}^2}]^2
\end{eqnarray}

The equations which determine $v$ and $\bar v$ then have the form

\begin{eqnarray}
\bar v( \alpha + \beta v \bar v) + 
Re[A^*( K + A v \bar v)] \bar v (v^2 + \bar v^2)
+ v ( K + A v \bar v)( K + A v \bar v)^* + \nonumber\\
v [ M_1^2 + g_1^2 (v^2 - {\bar v}^2)] + 
v\frac{g_2^2}{2} [ 2(v^2 - \bar v ^2) + {\kappa_u }^2 - {\kappa_d}^2] 
&=& 0
\end{eqnarray}

\begin{eqnarray}
v(\alpha + \beta v \bar v ) + 
Re[A^*(K + A v \bar v)] v ( v^2 + \bar v ^2 )
+ \bar v ( K + A v \bar v )( K + A v \bar v )^* \nonumber\\
+ \bar v [ M_2^2 - g_1^2 (v^2 - \bar v ^2)]
-\bar {v}\frac{g_2^2}{2}[ 2(v^2 - \bar v^2) + 
{\kappa_u}^2 - {\kappa_d}^2]
&=& 0
\end{eqnarray}

Defining
 
\begin{eqnarray}
M^2 &=& (M_1^2 + M_2^2)/2 \\
\delta &=& (M_1^2 - M_2^2)/2 \\
\chi &=& \delta + g_1^2[ v^2 - \bar v^2]
+g_2^2[ 2(v^2 - \bar v^2) + (\kappa_u)^2 - (\kappa_d)^2]/2 \\
\Lambda &=& \delta + g_1^2[ v^2 - \bar v^2] 
-g_2^2[ 2(v^2 - \bar v^2) + (\kappa_u)^2 - (\kappa_d)^2]/2
\end{eqnarray}
Let us keep in mind that the parameters $M^2$, $\delta$, $\chi$ and 
$\Lambda$ are all
of order $M^2_{SUSY}$, the mass scale for the soft terms; $A$ and $B$ are
of order $1/M_{Pl}$ and the  $\alpha$-term depends on the 
vev's of the singlet fields and could therefore be arbitrary.

Let us now rewrite the extremization equations as

\begin{eqnarray}
(\alpha + \beta v \bar v) + Re[A^*(v^2 + \bar v^2)(K + A v \bar v)]
\nonumber\\
+ (K + A v \bar v)(K + A v \bar v)^* + M^2 
&=& -\chi (v - \bar v)/(v + \bar v)
\end{eqnarray}

\begin{eqnarray}
-(\alpha + \beta v \bar v) - Re[A^*(K + A v \bar v)]( v^2 + \bar v^2)
\nonumber\\
+ (K + A v \bar v)(K + A v \bar v)^* + M^2 
&=& -\chi (v + \bar v)/(v - \bar v)
\end{eqnarray}
Let us note that $K+Av\bar v$ and $\alpha +\beta v\bar v$ vanish in the 
SUSY limit. If we assume that
the only source of supersymmetry breaking is from the soft breaking terms  
and if none of the singlets has vevs far exceeding the right handed scale
$M_R$ then careful
analysis shows that these are generically at most of order $M_{SUSY}$
and $M^2_{SUSY}$ respectively.
 This provides a qualitative way to see why the masses of the 
doubly charged fields are small compared to the $SU(2)_R$ scale since
it is these combinations that appear in the mass matrix for the 
doubly charged particles.       

To prove our result in more detail, let us multiply the above two 
equations to get a result which will be useful in the subsequent discussion:

\begin{eqnarray}
[ (K + A v \bar v)(K + A v \bar v)^* + M^2 ]^2 - 
[ ( \alpha + \beta v \bar v) + Re[A^*( K + A v \bar v)]
(v^2 + \bar v^2)]^2 &=& \chi^2
\end{eqnarray}

We now calculate the mass matrix for the doubly charged Higgs bosons and
obtain it to be

\begin{eqnarray}
\bordermatrix{ & {\Delta^c}^{--} & {\overline {\Delta}^c}^{--} \cr
{\Delta^c}^{++} & [K+(A+2B)v\bar v][h.c] + M^2 + \Lambda
& (\alpha_0^*+ \beta v \bar v) + 
(v^2 + \bar v ^2 )(A+2B)^*( K + A v \bar v ) \cr
{\overline {\Delta}^c}^{++} & (\alpha_0 + \beta v \bar v) + 
(v^2 + \bar v ^2 )(A+2B)( K + A v \bar v )^* &
[K+(A+2B)v\bar v][h.c] + M^2 - \Lambda \cr }
\end{eqnarray}

If either of the eigenvalues of this matrix is negative then 
the square of one of the scalar masses is negative and our
assumption that the theory preserves electric charge is invalid. Rather
than calculate the eigenvalues directly we choose to infer information
by examining the trace, T and determinant, D of the above matrix. If 
either of
these turns out to be negative the theory breaks electric charge.We first
determine the trace which is the sum of the eigenvalues as

\begin{eqnarray}
T &=& 2[K + (A + 2B)v \bar v][h.c.] + 2 M^2
\end{eqnarray}

This is typically of order O($M_{SUSY}^2$) or O($M_R^4/M^2_{Planck}$) where
 $M_{SUSY}$ is  
the scale of the soft SUSY breaking mass terms and $M_R$ the right handed
scale. Since the
product of the eigenvalues is merely the determinant we proceed to evaluate
this.

\begin{eqnarray}
D &=&
[(K + (A + 2 B) v \bar v )(h.c) + M^2]^2 \nonumber\\
&&- [(\alpha + \beta v \bar v) + 
(v^2 + \bar v ^2 )Re[( A + 2 B )( K + A v \bar v)^*]]^2 \nonumber\\
&&- [\alpha' + Im [(A +2B)(K+Av\bar v)^*](v^2+\bar v^2)]^2 - \Lambda ^2 
\end{eqnarray}

Using Eq.(10) and (20), this simplifies to
 
\begin{eqnarray}
D &=& [ \chi ^2 - \Lambda ^2 ] + \nonumber\\
&& (4 Re[B v \bar v ( K + A v \bar v)^*] + 4 B B^* v^2 \bar v^2)^2 \nonumber\\
&& -  (v^2 + \bar v^2 )^2[(2 B )(K + A v \bar v )][h.c.] \nonumber\\
&& + 8 (Re[(B v \bar v)( K + A v \bar v)^*] + B B^* v^2 \bar v^2)
[(K + A v \bar v )(h.c.) + M^2 ] \nonumber\\
&& - 4 [\alpha + \beta v \bar v + Re[A^*(K + A v \bar v)](v^2 + \bar v^2)]
[Re[B^*(K + A v \bar v)](v^2 + \bar v^2)]
\end{eqnarray}

We now have enough information to estimate the masses of the Higgs bosons.
For simplicity we subdivide our analysis into two cases:

1) low $W_R$ scale, so that all terms suppressed by powers of $ M_{Planck}$
   can be neglected; 
   
2) high $W_R$ scale.

We first consider the low $W_R$ scale case. This then corresponds to the
renormalizable theory, ie $A = B=0$. The only term present is 
$ \chi^2 - \Lambda^2 $
which we now examine in more detail. Using 

\begin{equation}
\Lambda = \chi - g_2^2[ 2( v^2 -\bar v ^2 ) + {\kappa_u}^2 - {\kappa_d}^2]
\end{equation}

and

\begin{equation}
\chi = - (v^2 - \bar v^2)(K^2 + M^2 )/(v^2 + \bar v ^2) 
\end{equation}
We find,
\begin{eqnarray}
\chi ^2 - \Lambda ^2 &=& -g_2^4 [ 2(v^2 -\bar v ^2)  + 
{\kappa_u}^2 - {\kappa_d}^2 ]^2 \nonumber\\
&& -4 g_2^2(v^2 - \bar v^2)^2[ K^2 + M^2]/(v^2 + \bar v ^2)
\nonumber\\
&& -2g_2^2(v^2 - \bar v^2)[K^2 + M^2]
({\kappa_u}^2 - {\kappa_d}^2)/(v^2 + \bar v ^2) 
\end{eqnarray}

In the limit that electroweak effects are ignored (i.e. 
$\kappa_u=\kappa_d=0$) this is less than zero
reproducing the known result\cite{kuchi} that the renormalizable theory 
has no charge
conserving vacuum. We now see however that the last term in (27) may in
fact alter this result if 
$|-{\kappa_u}^2 + {\kappa_d}^2|>2 |v^2 - \bar v ^2|$ 
and the $W_R$ scale is low. We can estimate the mass of the lightest
doubly charged boson to be $M_{++}\leq 
\frac{1}{4}\frac{(|\kappa^2_u-\kappa^2_d|)}{\sqrt{v^2+\bar{v}^2}}$.
This implies that unless the scale of right-handed symmetry breaking
(i.e. $\sqrt{v^2+\bar{v}^2}$) is less than about 400 GeV. For the case of
manifest left-right symmetry, such a low value for $M_{W_R}$ is inconsistent
with observations. Thus for the R-parity conserving vacuum, low $M_{W_R}$
scenario is inconsistent.

We now consider the high $W_R$ case. Now however, the theory with the
non-renormalizable operators can lie in the charge preserving vacuum for
sufficiently high $W_R$ scale. We estimate
the mass of the lighter of the doubly charged particles is either
O($M_R \sqrt{M_{SUSY}/M_{Planck}})$ or O($M_R^2/M_{Planck}$), whichever is 
larger, while the mass of the heavier is the larger of O($M_R^2/M_{Planck}$)
and O($M_{SUSY}$). The experimental lower bound on the mass of the lighter
 particle implies that $M_R > 10^9$ GeV.

In both the low and high $W_R$ cases as a consequence of supersymmetry breaking
 the Higgsinos pick up a mass from the following term in the superpotential:

\begin{equation}
W = [ K + A v \bar v +2 B v \bar v ] 
{\Delta^c}^{--}{\overline{\Delta}^c}^{++} \end{equation}

This is a Dirac mass of order O$(M_{SUSY})$ or $M^2_R/M_{Pl}$, whichever
is larger.

However in view of the fact that the high $W_R$ scenario necessarily 
envisages a
large hierarchy between the right handed scale and the scale of the soft
SUSY masses, an effective field theory calculation would be more convincing
than our tree level result. For simplicity we now restrict ourselves to the
case of a single singlet $S$, which we integrate out at tree level along 
with the other heavy fields ${\Delta^c}^0$, ${\overline{\Delta}^c}^0$
 ,${{\Delta}^c}^-$,
and ${\overline{\Delta}^c}^+$. The remaining effective field theory, 
consisting
of ${\Delta^c}^{--}$, ${\overline{\Delta}^c}^{++}$ and some of the $X_i$'s 
is then
run down to the scale of the light Higgs fields. The potential we start
 from is the same as before except that $ G(S_i,X_i)$ and 
 $G'[S_i,X_i,S_i^{\dagger},X_i^{\dagger}]$ become
$ G(S,X_i)$ and $G'[S,X_i,S_i^{\dagger},X_i^{\dagger}]$ respectively while
 $ \lambda_i$ and $\lambda'$ are now simply $ \lambda $ and $\lambda'$.

We expand ${{\Delta^c}^0}$, ${\overline {\Delta^c}}^0$, and $S$ about their 
vacuum expectation values as 

\begin{eqnarray} 
{\Delta^c}^0 = v + \eta \\
{\overline{\Delta}^c}^0 = \bar v +\bar {\eta} \\ 
S = < S_0 > + S' 
\end{eqnarray}

Defining $\sigma_1= v  \bar{\eta} + \bar v  \eta$ and 
$\sigma_2$ =$v$ Re $\eta - \bar v$ Re $\bar{ \eta}$, we
find that to the extent that supersymmetry breaking terms
and terms suppressed by powers of $M_{Planck}$ are small, 
$\sigma_2$ and two linearly independent combinations of 
$\sigma_1$ and $S$ are approximate mass eigenstates. We define these 
two linearly independent combinations as $S_1$ and $S_2$. 
Then it is straightforward to verify that as a consequence
of the cancellations of the tree level graphs involving the exchange
of $\sigma_1, S_1$ and $S_2$, 
to order O($M_{SUSY}/M_R$) or O($M_R/M_{Planck}$) the
only residual interactions among the light fields are those
in the effective potential below. We can write the part of the effective
potential relevant for the light doubly charged Higgs field
as 

\begin{equation}
V = ( [\int d^2 \theta W + h.c.] + V_S + V_D )
\end{equation}

where

\begin{equation}
W = [ K + A v \bar v +2 B v \bar v ] 
{\Delta^c}^{--}{\overline{\Delta}^c}^{++}
    +fe^ce^c{\Delta^c}^{--}
\end{equation}
     
\begin{eqnarray}
V_{soft} &=& [( \alpha_0 + \beta v \bar v + (v^2 + \bar v ^2)(A + 2B)
           (K + A v \bar v)^*){\Delta^c}^{--}{\overline{\Delta}^c}^{++} + 
h.c.]
           \nonumber\\&& + ( M^2 + \Lambda){\Delta^c}^{++}{\Delta^c}^{--} 
           + ( M^2 - \Lambda){\overline{\Delta}^c}^{++}
           {\overline{\Delta}^c}^{--}
           +f'M''\tilde{e^c}\tilde{e^c}{\Delta^c}^{--}+h.c.
\end{eqnarray}

\begin{eqnarray}
V_D &=&\frac{2g^2_1 g^2_2}{(g^2_1+g^2_2)}[{\Delta^c}^{++}{\Delta^c}^{--} - 
{\overline {\Delta}^c}^{--}{\overline{\Delta} ^c}^{++}]^2 
\end{eqnarray}
                   
This effective theory must now be run down to the mass scale of the 
doubly charged fields $M_{\Delta}$ using the renormalization group 
equations ( R.G.E.'s). On writing
down the relevant R.G.E.'s it is clear that the only potentially
large contribution 
to the evolution of the boson mass terms is likely to arise from the 
coupling to the leptons f and f' and is of order 
O ($ \sqrt {f^2 M_{SUSY}^2 ln [ M_R/ M_{\Delta}]/ 8 \pi^2} $).
 Hence our tree level result
for the mass terms may be corrected by about this amount. Clearly
this will not qualitatively alter our result that the doubly charged
Higgs bosons will be light. The fermion masses suffer only wave 
function renormalization and also remain light.             
          
\section{Analysis for $W_R$ Scale above SUSY breaking Scale}

In the previous section we assumed that the scale at which 
supersymmetry is broken is higher than the $W_R$ scale. However this
need not be the case and in particular there has recently been a lot
of interest in theories where gauge interactions are the mediators 
of supersymmetry breaking at a relatively low scale\cite{dine}. This is the 
case we now study in detail.

Our analysis essentially will differ from that of the previous section
only in that the soft SUSY breaking terms are now generated explicitly 
only at 
the scale at which the messenger fields are integrated out, and are
not explicitly present at the $W_R$ scale. Since they are generated by 
loop graphs involving the gauge bosons of the residual symmetries,
their form will be such as to respect only the surviving gauge symmetries.
 We will show that this difference 
has consequences for phenomenology. Our procedure must therefore be 
to integrate out the heavy fields at the $W_R$ scale, run the theory
down to the messenger scale, integrate out the messengers thereby 
generating the soft SUSY breaking mass terms, and then run the theory 
down to the $M_{\Delta}$ scale. We make no assumption about the 
messenger fields except that they carry electroweak quantum numbers,and
do not couple directly to the Higgs sector.
However for simplicity we restrict ourselves once again to the one 
singlet case.

After integrating out the heavy fields at the right handed scale,
the effective field theory has the form

\begin{equation}
V = [ \int d^2 \theta W + h.c. ] + V_D
\end{equation}

where $V_D$ is the same as in Eq.(5) but $W$ is now simply

\begin{eqnarray}
W = 2 B v \bar v {\Delta^c}^{--}{\overline{\Delta}^c}^{++}
    +fe^ce^c{\Delta^c}^{--}
\end{eqnarray}

This potential suffers only wave function renormalization down to the 
messenger scale. Then on integrating out the messenger fields soft
SUSY breaking terms will be generated, the form of which are to some
extent dependent on the nature of the messengers. However, if the
messengers couple to the Higgs sector only through gauge interactions
and not directly through the superpotential then the relevant part of
these terms generically have the form,

\begin{eqnarray}
V_S &=& M^2( {\Delta^c}^{++}{\Delta^c}^{--}+{\overline {\Delta}^c}^{--} 
{\overline{\Delta}^c}^{++})
\end{eqnarray}

These terms arise from two loop diagrams involving the messenger fields
coupling via hypercharge gauge interactions to the Higgs sector. 
The total potential must then be run down to the electroweak scale
using the same R.G.E.'s as in the previous section, and the mass term
will once again receive some modification. However, this cannot alter
the basic result that the doubly charged Higgs bosons only acquire a
mass of order O($M_{SUSY}$) or O($M_R^2/ M_{Planck}$) and therefore 
remain light. Our analysis however brings up the following interesting
question; in such a scenario can the theory live in the charge 
preserving vacuum without the need for higher dimensional operators?
After all, since the light fields all have positive mass even in the 
absence of the higher dimensional operators the charge conserving 
vacuum is at least a local vacuum of the theory even without them!
Notice however that the doubly charged Higgsinos then have no mass.
Thus the nonrenormalizable terms are still required, this time to 
give mass to the Higgsinos. The experimental lower bound on the mass 
of such a doubly charged fermion puts a bound on the $W_R$ scale
of $M_{W_R} \geq 10^{10}$ GeV. 

It is interesting that the theory can in fact live in the charge 
preserving vacuum without the higher dimensional operators. The
possibility exists that by bringing the $W_R$ and SUSY breaking scales
close together or by reconsidering the assumption that the messenger 
and Higgs sectors do not directly couple in the superpotential it may
be possible to do away with the higher dimensional operators 
altogether, thereby altering the bound. This is a possible direction for 
future research.

\section{Models with Additional Matter Content}

In this section we consider the effects of relaxing our earlier
restrictions on the matter content of the model we studied on our
result that the doubly charged Higgs bosons and Higgsinos are light.
 In order
for the additional matter content to affect our result there must be
 a difference between the
vacuum energies of the charge conserving and charge violating vacua or
at least a barrier between them. This must occur in the limit of
exact supersymmetry because any correction to the masses from SUSY
breaking effects will be at most of order $M_{SUSY}$ and too small to
fundamentally alter our result.We also ignore nonrenormalizable terms
because corrections to the masses arising from these can reasonably be
expected to be small.Hence we will be studying renormalizable theories 
in the limit of exact supersymmetry and observing the effect of the 
additional matter on the masses of the doubly charged Higgs fields.We
also assume that R-parity is unbroken.

We now attempt to systematically go over some possibilities for the 
additional matter content. Since the new fields transform as 
representations of $SU(2)_R$ we proceed in order of increasing 
dimensionality of the representation.

\noindent 1) Charged singlets $T(1, 4)$ and $\overline T(1, -4)$

Here the numbers within the parenthesis denote the $SU(2)_R$ and
$U(1)_{B-L}$ quantum numbers respectively. These allow 
couplings of the form $ T~ Tr(\Delta^c \Delta^c) $ and 
$ \overline T~Tr(\overline \Delta^c \overline \Delta^c)$ in the
superpotential. It is straightforward to verify that these do indeed
lift the flat direction giving the components of the doubly charged 
Higgs superfield a mass at the $W_R$ scale.

\noindent 2) Charged vector triplets $T(3,2)$ and $\overline T(3,-2)$
 
These 
have the same quantum numbers as $\Delta^c $ and $\overline \Delta^c$
Hence superpotential couplings of the form $M Tr (\Delta^c T)$ ,
$ M Tr (\overline T ^{ } \overline \Delta^c)$, $S_i Tr (\Delta^c T)$ and 
$ S_i Tr (\overline T ^{ } \overline \Delta^c)$ are now possible.However 
these will not lift the flat direction because the vevs of $T$ and 
$\overline T$ can always have the same form as those of 
$\overline \Delta^c$ and $\Delta^c $ without any effect of the 
angle $\theta$ on the vacuum energy ie if 
       
\begin{eqnarray}
\Delta^c =v 
\left[
\matrix{ {0} & { cos  \theta} \cr
         { sin \theta} & {0} \cr}
\right]
\end{eqnarray}  

then

\begin{eqnarray}
\overline T =t 
\left[
\matrix{ {0} & { cos  \theta} \cr
         { sin \theta} & {0} \cr}
\right]
\end{eqnarray}  

Thus we do not expect our result to change.

Apart from the charged singlet above, no field transforming as a 
higher dimensional representation of $SU(2)_R$ when added to the
theory can preserve the charge conserving vacuum unless it itself
also breaks $SU(2)_R$. This may be verified by explicit calculation
for all the three cases below which exhaust the possibilities.
This is because supersymmetry
is explicitly broken by the superpotential in the charge preserving
vacuum unless the additional matter fields pick up vevs thereby 
breaking $SU(2)_R$ themselves.

\noindent 3) A neutral triplet\cite{goran2} $\Omega^c(3,0)$ 
that couples to 
the Higgs sector as $Tr (\Omega^c {\overline {\Delta}^c}\Delta^c)$

\noindent 4) A neutral quintuplet $T_{ij}^{kl}$(5,0) that couples as 
${{\Delta}^c}^i_k {\overline {\Delta}^c}^j_l T_{ij}^{kl}$.

\noindent 5) Charged quintuplets $T_{ij}^{kl}$(5, 4) and 
$\overline T_{ij}^{kl}$(5, -4) that couple to the Higgs sector as 
${{\Delta}^c}^i_k {{\Delta}^c}^j_l T_{ij}^{kl}$ and
${\overline {\Delta}^c}^i_k {\overline {\Delta}^c}^j_l
\overline T_{ij}^{kl}$.

If the multiplets that have been added to the theory break $SU(2)_R$
then only a detailed analysis of the vacuum structure for each 
individual theory can determine whether the relevant flat direction
is lifted or not. There are also far more possibilities than the 
three above. A careful analysis has been performed for the neutral 
triplet\cite{goran2}, which shows that the flat direction is successfully 
lifted, but not for the other cases. To do so for the other cases
is beyond the scope of the present paper.

Thus the conclusion of this section is that the light doubly
charged Higgs superfield can be avoided if a doubly charged $SU(2)_R$
singlet is present in the theory or if there are certain specific
Higgs multiplets which break $SU(2)_R$ while preserving hypercharge. 

\section{Conclusion}

     In summary, we find that the combination of supersymmetry with
left-right symmetry leads to nontrivial constraints on the mass spectrum 
of the theory if the starting theory is assumed to be automatically R-parity
conserving. In particular, we find the interesting result that for the 
R-parity conserving scenario, the 
mass of the doubly charged Higgs bosons and/or Higgsinos which are part
of the $SU(2)_R$ multiplets used to implement the see-saw mechanism
will be unacceptably light unless the $W_R$ mass is larger
than $10^{9}$ GeV. Thus the only situation where the low scale of parity 
restoration and supersymmetry are consistent with each other is when 
R-parity is dynamically broken by the vacuum. 

Our result has the following interesting implications:

\noindent (i) This should give new impetus to the
experimental searches for the $W_R$ boson, since it implies that if 
experiments exclude a low mass ${W_R}$, then its mass can 
only be in the $10^{9}$ GeV.
This latter range is of course of great deal of interest in connection
with solutions to the solar and atmospheric neutrino puzzles. On the other
hand, if a low mass $W_R$ is discovered, it would imply that in the context
of simple models that R-parity must be dynamically broken.

\noindent (ii) The lightness of the doubly charged fields
is now valid even if the $SU(2)_R$ scale is in the superheavy 
range (i.e. $>10^{9}-10^{10}$ GeV); discovering the phenomenological
effects of these light particles\cite{huitu} 
acquires a new urgency and importance.
The experimental discovery of such particles would provide spectacular
evidence for the supersymmetric left-right model and their masses 
could provide valuable information about the $W_R$ scale.

\noindent (iii) Our result will have important implications for gauge
and Yukawa coupling unification in left-right and SO(10) models with
automatic R-parity conservation. In particular, the evolution equations
will have to include the effects of the doubly charged particles at 
a much earlier scale than the $SU(2)_R$ breaking scale. Otherwise
the exotic multiplets that help us to avoid our result will have
to be included above the $W_R$ scale. This is presently under investigation.

\noindent (iv) Finally, our results about the lightness of the doubly charged
Higgs bosons hold even when R-parity is spontaneously broken, since as
already emphasized, in this case the $SU(2)_R$ scale is bound to be in the
TeV range\cite{kuchi}.

\noindent{\bf Acknowledgement}

One of us (Z. C.) would like to thank Markus Luty for discussions. 
This work has been supported by the National Science Foundation grant
number PHY-9421386.

\end{document}